\RequirePackage{lineno}
\documentclass[article,prl,
twocolumn,
amsmath,amssymb,
amsmath,amssymb, aps,
]{revtex4}

\usepackage[dvipsnames]{xcolor} 
\usepackage{epsfig}
\usepackage{amsfonts}
\usepackage{amsmath}
\usepackage{wasysym}
\usepackage{amssymb}
\usepackage{bm} 
\usepackage[framemethod=PStricks]{mdframed}
\usepackage{lipsum}
\usepackage{float}

\usepackage{pdflscape}
\usepackage{graphicx}
\usepackage{ulem}

\newcommand{\be}{\begin{equation}}
\newcommand{\ee}{\end{equation}} 
\newcommand{\bea}{\begin{eqnarray}} 
\newcommand{\eea}{\end{eqnarray}}
\newcommand{\bfr}{{\bf{r}}}
\newcommand{\bfk}{{\bf{k}}}

\usepackage[utf8]{inputenc}
\usepackage{bbm} 				
\usepackage{epstopdf}				
\usepackage{verbatim}    			
\usepackage{sidecap}                
\usepackage{indentfirst}
\usepackage[colorinlistoftodos]{todonotes}

\begin{document}

\title{Vortex Gas Modeling of Turbulent Circulation Statistics}

\author{G.B. Apolinário$^{1}$, L. Moriconi$^{1}$, R.M. Pereira$^{2}$, and V.J. Valadão$^{1}$}
\affiliation{$^{1}$Instituto de F\'\i sica, Universidade Federal do Rio de Janeiro, \\
C.P. 68528, CEP: 21945-970, Rio de Janeiro, RJ, Brazil}
\affiliation{$^{2}$Instituto de Física, Universidade Federal Fluminense, 24210-346 Niterói, RJ, Brazil}


\begin{abstract}
Statistical properties of circulation encode relevant information about the 
multi-scale structure of turbulent cascades. Recent massive computational 
efforts have posed challenging theoretical issues, as the dependence of circulation 
moments upon Reynolds numbers and length scales, and the specific shape of 
the heavy-tailed circulation probability distribution functions. 
We address these focal points in an investigation of circulation statistics 
for planar cuts of three-dimensional flows. The model introduced here borrows 
ideas from the structural approach to turbulence, whereby turbulent flows are 
depicted as dilute vortex gases, combined with the standard Obukhov-Kolmogorov 
phenomenological framework of small-scale intermittency. We are able to 
reproduce, in this way, key statistical features of circulation, in close 
agreement with empirical observations compiled from direct numerical 
simulations.
\end{abstract}


\maketitle



Drawing analogies with the Wilson loop strategy to tackle the quark confinement problem \cite{wilson,migdal1},
Migdal introduced, some 25 years ago, alternative circulation functional methods to the context of 
fully developed turbulence \cite{migdal2}. The subject of turbulent circulation has now been vigorously revived both on the theoretical and numerical fronts. 
Interesting ideas have reached firmer grounds, as the area law for the probability distribution function of circulation  \cite{migdal3,Iyer_etal,migdal4}, whereas unexpected phenomena have 
been additionally discovered, as the bifractal scaling behavior of circulation moments \cite{Iyer_etal}.

In this work, focused on the problem of isotropic and homogeneous turbulence, we define the circulation 
variable simply as
\be
\Gamma_R \equiv \int_\mathcal{D} d^2 \bfr \, \omega(\bfr)  \ , \ \label{circ} 
\ee
where $\mathcal{D}$ is a circular domain of radius $R$ which lies in a plane $\gamma$ and $\omega(\bfr)$ 
is the component of the vorticity field which is normal to $\gamma$ (an arbitrary orientation is
chosen). Our aim is to explore the sensitivity of (\ref{circ}) to the 
presence of vortex tubes -- so clearly identified in turbulent flows since the early 1990's \cite{orszag_etal, farge_etal, kaneda_etal} --
as a natural modeling perspective to account for relevant empirical findings.

Kolmogorov's phenomenological description of turbulence (K41) \cite{K41} suggests that if $R$ is far from the energy injection and the Kolmogorov dissipative scales, $L$ and $\eta_K$, respectively, the only other relevant physical parameter related to the statistical behavior of velocity fluctuations at scale $R$ is the energy dissipation rate per unit mass, $\epsilon$. We define, as usual, 
$\eta_K = (\nu^3 / \epsilon)^{1/4}$, where $\nu$ is the kinematic viscosity of the fluid. 
Straightforward dimensional analysis leads, then, to circulation moments {\hbox{$\langle \Gamma_R^p \rangle \sim \epsilon^{\frac{p}{3}} R^{\frac{4p}{3}}$}}. Having in mind a reinterpretation of this scaling law within the structural context, where it is assumed that most of the turbulent kinetic energy is generated by vortex tubes \cite{farge_etal}, consider the second order moment of $\Gamma_R$, 
$
\langle \Gamma_R^2 \rangle = \int_\mathcal{D} d^2 \bfr \int_\mathcal{D} d^2 \bfr' \langle \omega(\bfr) \omega(\bfr') \rangle.
\label{Gamma2}
$
Taking this expression into account and noticing that the K41 prediction for order $p=2$ can be reshuffled as {{\hbox{$\langle \Gamma_R^2 \rangle \sim (R / \eta_K)^4 [ ( \epsilon / \nu )^{\frac{1}{2}} \eta_K^2 ]^2 (\eta_K / R )^{\frac{4}{3}}$}}}, one may suggest that circulation is effectively produced by (i) $N \propto (R / \eta_K)^2$ planar vortices which have (ii) rms vorticities of the order of $( \epsilon / \nu )^{\frac{1}{2}}$, (iii) core sizes of linear dimensions of the order of $\eta_K$, and (iv) carry elementary circulations which are correlated at separation distance $r$ as $\sim 1/r^{4/3}$, for $r \gg \eta_K$.

\begin{figure}[ht]
\includegraphics[width=0.4\textwidth]{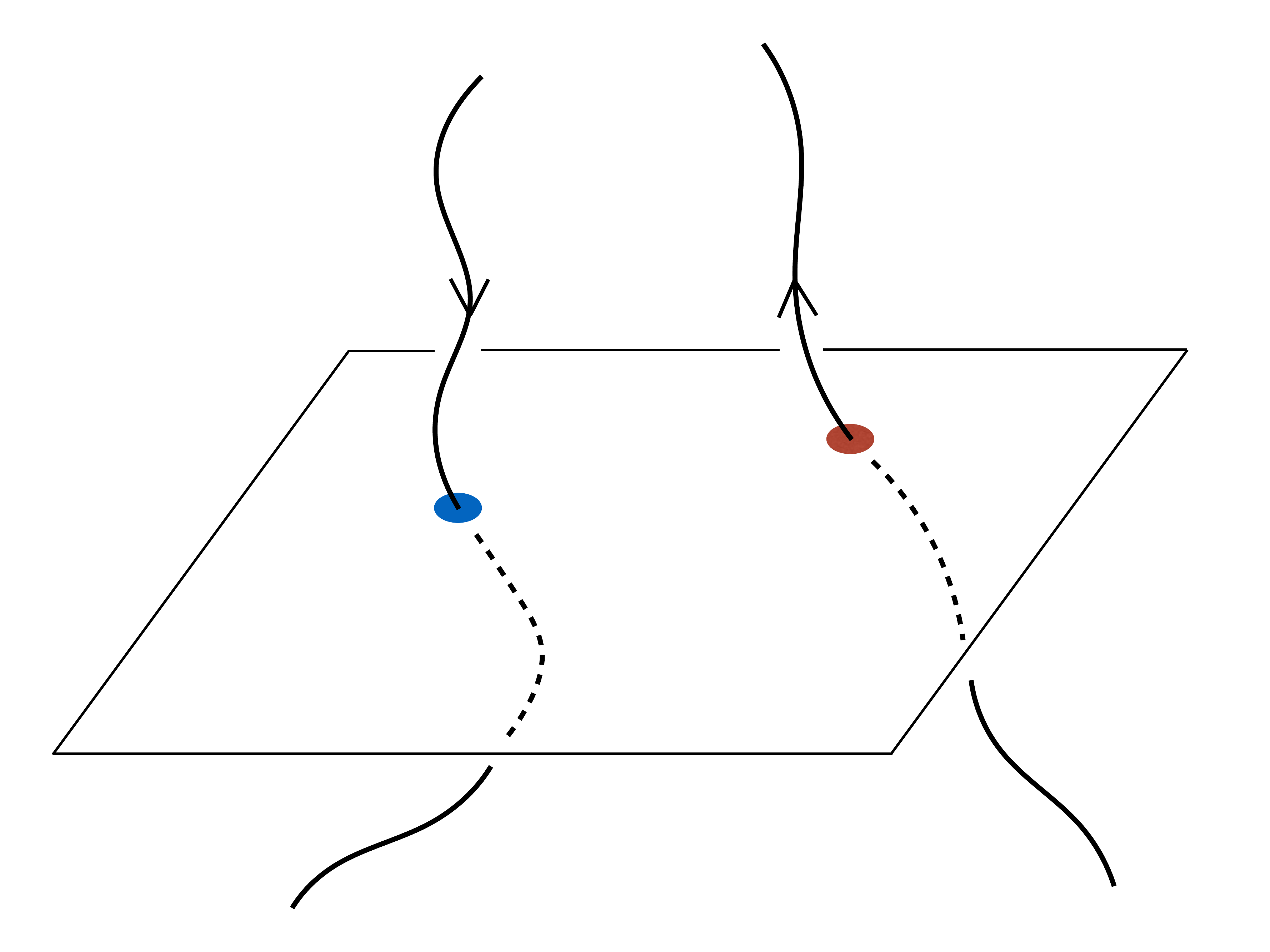}
\caption{Two thin vortex tubes with opposite orientations cross a plane and are associated, in this particular example, to a planar vortex-antivortex pair (red and blue spots, respectively) where vorticity is concentrated. In more general terms, 
for any given fixed configuration of planar vortices defined by means of a similar construction, there is a large statistical 
ensemble of three-dimensional vortex tubes that produce a (conditioned) random vorticity field on the plane.}
\label{vortices}
\end{figure}

Two important issues have to be dealt with, before proceeding along with this structural line of reasoning. The first issue is the 
apparent dichotomy between the acknowledged existence of three-dimensional vortex tubes and the introduction of planar vortices in the modeling definitions. A possible solution of this problem is to take the positions of the effective planar vortices as the intersections of vortex tubes with the plane $\gamma$ that contains the circular domain $\mathcal{D}$. Now, figuring out the large ensemble of three-dimensional flow configurations conditioned by a fixed spatial distribution of $N$ intersecting vortex tubes in $\gamma$ at positions $(\bfr_1, \bfr_2,...,\bfr_N)$, it is clear that the vorticity vector at position $\bfr$ in $\gamma$ can be represented as the conditioned random field
$\boldsymbol\omega (\bfr)  = \boldsymbol\omega (\bfr | \bfr_1, \bfr_2,...,\bfr_N)$,
which is negligible if probed far enough from the planar vortex spots, that is, at positions $\bfr$ such that $|\bfr - \bfr_i| \gg \eta_K$ for $i=1,2,...,N$. The physical picture addressed here is illustrated in Fig. \ref{vortices}.

Relying upon the fact that $\boldsymbol\omega (\bfr)$ is the superposition of the vorticity fields produced by a random system of three-dimensional vortex tubes, it is tempting to evoke the central limit theorem in some of its functional generalizations \cite{bolt}, to take $\boldsymbol\omega (\bfr)$ as a Gaussian random field. This is actually a meaningful hint, but some care is necessary on this point - and here it comes a second modeling difficulty: we have to include intermittent fluctuations of the energy dissipation rate in our arguments. In the Obukhov-Kolmogorov phenomenology of intermittency (OK62) \cite{O62,K62} and its subsequent developments \cite{frisch,ro_va}, local dissipation is described, to very good approximation, as a lognormal, long-ranged correlated field. Insisting in the modeling points (ii), (iii) and (iv), introduced above, we collect all the pieces of information brought to the discussion so far, to write
\be
\omega(\bfr) \propto \sum_{i=1}^N g_\eta(\bfr - \bfr_i) \xi (\bfr_i) \tilde \omega (\bfr_i) \ . \
\label{omega_sum}
\ee
Here $g_\eta(\bfr) \equiv \exp [ - \bfr^2 / (2 \eta^2)  ]$ introduces Gaussian envelopes for the planar vortices of typical width $\eta$ which, in consonance with \cite{jimenezetal93}, are taken proportional to $\eta_K$, i.e., $\eta = a \eta_K$ with $a$ being a positive modeling parameter. The field $\tilde \omega (\bfr)$ is a scalar Gaussian random field with vanishing mean and correlator 
$\langle \tilde \omega (\bfr) \tilde \omega (\bfr') \rangle  \sim 1/|\bfr - \bfr'|^{4/3}$ 
for $|\bfr - \bfr'| \gg \eta_K$, while
$
\xi(\bfr) \equiv \xi_0 \sqrt{\epsilon (\bfr) / \epsilon_0},
$
where $\xi_0$ is an additional positive parameter and $\epsilon (\bfr)$ is the energy dissipation rate at position $\bfr$, which has mean value $\epsilon_0 = \langle \epsilon (\bfr) \rangle$, to be modeled within the OK62 framework. We call attention, in connection with Eq. (\ref{omega_sum}), to the general fact that the statistical dependence of velocity gradients with the square root of the dissipation field is the usual way to extend the OK62 description of intermittency to the dissipative scale region \cite{wyn_tenne,kholmy_etal,pereira_etal}.
We take $g_\eta(\bfr)$, $\xi(\bfr)$ and $\tilde \omega (\bfr)$ to be dimensionless quantities. Without loss of generality, the variance of $\tilde \omega (\bfr)$ is prescribed to unity.

The positions and the number $N$ of vortices in (\ref{omega_sum}) are of course random quantities that depend on the shape and the size of the planar domain crossed by the vortex tubes. The simplest modeling choice here is to assume that vortices are randomly distributed over $\gamma$ with a Poissonian surface density whose mean we denote as $\bar \sigma \equiv \langle \sigma(\bfr) \rangle$. Introducing a convenient prefactor with dimensions of vorticity, we write, motivated by (\ref{omega_sum}),
\be
\omega(\bfr) = \sqrt{\frac{\epsilon_0}{3 \nu}} \int d^2 \bfr' g_\eta (\bfr -\bfr')  \xi(\bfr') \tilde \omega (\bfr')  \sigma(\bfr') \ . \
\label{omega}
\ee
Note, as well-known from the statistical theory of turbulence \cite{batchelor}, that due to isotropy we expect to have {\hbox{$\langle \omega(\bfr)^2 \rangle = \langle \boldsymbol\omega(\bfr)^2 \rangle/3 = \epsilon_0/(3 \nu)$}}. Furthermore, once direct numerical simulations indicate that the volume occupied by vortex tubes is a very small fraction of the total fluid volume \cite{farge_etal}, we may employ a dilute vortex gas approximation to represent density fluctuations as $\sigma (\bfr) = \bar \sigma + \phi(\bfr) $, with $ \langle \phi(\bfr ) \rangle =0$ and, up to fourth-order,
\bea
&& \langle \phi(\bfr_1 ) \phi(\bfr_2) \rangle = \bar \sigma \delta_{12} \ , \  \label{p2} \\
&& \langle \phi(\bfr_1 ) \phi(\bfr_2) \phi(\bfr_3) \rangle = \bar \sigma \delta_{12} \delta_{13} \ , \ \\ \label{p3}
&&\langle \phi(\bfr_1 ) \phi(\bfr_2) \phi(\bfr_3) \phi(\bfr_4) \rangle = \bar \sigma \delta_{12}\delta_{13}\delta_{14} + \nonumber \\
&&+ ~ \bar \sigma^2 ( \delta_{12} \delta_ {34} + \delta_{13} \delta_ {24} + \delta_{14} \delta_ {23} ) \ , \ \label{p4}
\eea
where we have used the notation $\delta_{ij} = \delta^2(\bfr_i - \bfr_j)$ \cite{comment}.

A smooth cutoff-regularized expression for the Gaussian random field $\tilde \omega (\bfr)$ proves to be of great analytical help in the evaluation of vorticity correlation functions. We just mean that $\tilde \omega (\bfr)$ is given as
\be
\tilde \omega ( \bfr ) = \frac{\eta_K^{\frac{2}{3}}}{\sqrt{2 \pi \Gamma \left (\frac{4}{3} \right )}} \int d^2 \bfk \, \psi (\bfk) k^{-\frac{1}{3}} \exp \left ( i \bfk \cdot \bfr - k \frac{\eta_K}{2} \right ) \ , \ \label{fbm}
\ee
where $\psi (\bfk )$ is a complex Gaussian random field of zero mean and
correlator
$
\langle \psi (\bfk_1 ) \psi (\bfk_2) \rangle = \delta^2(\bfk_1 -\bfk_2)
$.

The formulation addressed by Eq. (\ref{omega}) becomes in fact cumbersome if one is interested to compute moments of order $p>2$ for the circulation variable. The reason is that $\xi(\bfr)$ is not a Gaussian random field. A pragmatic phenomenological solution of this problem comes from an alternative way of performing the circulation integral. Considering, initially, that $R \gg \eta_K$, we obtain, from Eqs. (\ref{circ}) and (\ref{omega}), the asymptotic approximation
\be
\Gamma_R = 2 \pi \eta^2 \sqrt{\frac{\epsilon_0}{3 \nu}}  \int_\mathcal{D} d^2 \bfr \, \xi(\bfr) \tilde \omega (\bfr)  \sigma(\bfr) \ . \ \label{circ2}
\ee
Since $\xi(\bfr)$ is a positive definite quantity, it is not difficult to show that there is necessarily a point {\hbox{$\bfr_0 \in \mathcal{D}$}} such that 
{\hbox{$
\Gamma_R = 2 \pi \eta^2  \sqrt{\epsilon_0 / (3 \nu)} \tilde \omega (\bfr_0)  \sigma(\bfr_0) \int_\mathcal{D} d^2 \bfr \, \xi(\bfr)
$}}.
Owing, now, to the fact that $\xi(\bfr)$ is long-range correlated and that the probability measures for the $\tilde \omega(\bfr)$ and $\sigma(\bfr)$ fields are translation invariant, we expect that $\bfr_0$ will randomly ``slide", with approximate uniform distribution over the domain $\mathcal{D}$, in the ensemble of flow configurations conditioned to fixed $\int_\mathcal{D} d^2 \bfr \xi(\bfr)$.
In this case, Eq. (\ref{circ2}) can be effectively replaced by
\be
\Gamma_R = 2 \pi \eta^2  \sqrt{\frac{\epsilon_0}{3 \nu}}   \xi_R \int_\mathcal{D} d^2 \bfr \, \tilde \omega (\bfr)  \sigma(\bfr) \ , \
\label{circ4}
\ee
where 
\be
\xi_R \equiv \frac{1}{\pi R^2} \int_\mathcal{D} d^2 \bfr \, \xi(\bfr) = \frac{\xi_0}{\pi R^2} 
\int_\mathcal{D} d^2 \bfr \, \sqrt{\frac{\epsilon(\bfr)}{\epsilon_0}} \ . \ \label{circ5}
\ee
In other words, in order to compute statistical properties of $\Gamma_R$, for $R \gg \eta_K$, we just need to deal with the much simpler random vorticity field
\be
\omega_R (\bfr) = 2 \pi \eta^2  \sqrt{\frac{\epsilon_0}{3 \nu}}   \xi_R  \tilde \omega (\bfr)  \sigma(\bfr) \ . \ \label{omegaR}
\ee
Incidentally, and fortunately, the very same expression as the above one is supposed to hold for the small scale region $R \ll \eta_K$, since the dissipation field is not expected to exhibit fast spatial variations within dissipative length scales. 

The spatially averaged field $\xi_R$ has here a statistical role similar to the one of the coarse-grained dissipation field introduced in the OK62 phenomenology \cite{O62,K62}. Putting forward considerations which are nothing more than a direct application of OK62 ideas, we can write that
\be
\xi_R = \xi_0 \exp ( - X_R )  \ , \ \label{xiX}
\ee
where $X_R$ is a Gaussian random variable with both mean and variance given by
\be
\bar X_R = \frac{3 \mu}{8} \ln \left [ \frac{R_\lambda}{\sqrt{15}} \left ( \frac{\eta_K}{b R + \eta_K} \right )^{\frac{2}{3}} \right ] \ , \  \label{Xxi}
\ee
where $R_\lambda$ is the Taylor-Reynolds number, $\mu = 0.17 \pm 0.01$ is the intermittency exponent \cite{tang_etal}, and $b>0$ is a modeling parameter. The specific $R$ dependent expression between parentheses in (\ref{Xxi}) is proposed as an interpolation that works correctly for the $R /\eta_K \ll 1$ 
and $R / \eta_K \gg 1$ asymptotic cases. Note that (\ref{xiX}) and (\ref{Xxi}) lead to
$\langle \xi_R^4 \rangle \sim \langle \epsilon_R^2 \rangle \sim R^{-\mu}$ for $R \gg \eta_K$, as expected. 

We are now ready to explore the model predictions given by Eqs. (\ref{circ}), (\ref{p2}-\ref{p4}), and (\ref{omegaR}-\ref{Xxi}), where the underlying $\xi(\bfr)$, $\tilde \omega (\bfr)$, and $\sigma(\bfr)$ fields are assumed to be statistically independent from each other. Defining $\bar R = R / \eta_K$, the first interesting quantities to compute are the circulation variance and kurtosis in the asymptotic limit $\bar R \ll 1$. It turns out that up to lowest order in $\bar R$ dependence,
\bea
&&\langle \Gamma_R^2 \rangle = \frac{\epsilon_0}{3 \nu} \xi_0^2 \bar \sigma \frac{\pi^3 \eta^6}{a^4} \bar R^4  \ , \ 
\label{varsmallR} \\
&& 
\frac{\langle \Gamma_R^4 \rangle}{\langle \Gamma_R^2 \rangle^2}
= \frac{3}{2}  \frac{1}{\bar \sigma \pi \eta^2} \left ( \frac{R_\lambda}{\sqrt{15}} \right )^{\frac{3 \mu}{2}} 
\left ( 1 - \frac{1}{4a^2} \bar R^2 \right ) \label{k4lowR}
\ . \
\label{kappa4}
\eea
Recalling that for $\bar R \ll 1$ one has $\langle \Gamma_R^2 \rangle \simeq \langle \omega^2 \rangle (\pi R^2)^2 = \epsilon_0 (\pi R^2)^2 /(3 \nu)$, we find, from (\ref{varsmallR}),
\be
\xi_0^2 = \frac{1}{\bar \sigma \pi \eta^2} \ . \ \label{xi0}
\ee
%
Also, from the data of the numerical simulations reported in \cite{Iyer_etal} we obtain that $\lim_{\bar R \rightarrow 0}  
\langle \Gamma_R^4 \rangle / \langle \Gamma_R^2 \rangle^2 \simeq C_4 R_\lambda^{\alpha_4}$, where $C_4 \simeq 1.16$
and $\alpha_4 \simeq 0.41$. This implies, using (\ref{kappa4}), that
\be
\bar \sigma \pi \eta^2 = \frac{3}{2} \frac{1}{C_4} \frac{1}{15^{\frac{3 \mu}{4}}} R_\lambda^{\frac{3 \mu}{2} - \alpha_4} \ . \ \label{barsigma}
\ee
Since $3 \mu / 2  - \alpha_4 = -0.14 < 0$, the above result tells us that the dilute vortex gas approximation improves the higher is the Reynolds number. For $R_\lambda = 240$, for instance, we already have the satisfying estimate
$\bar \sigma \pi \eta^2 = 0.39$. 

\begin{figure}[ht]
\hspace{0.0cm} \includegraphics[width=0.48\textwidth]{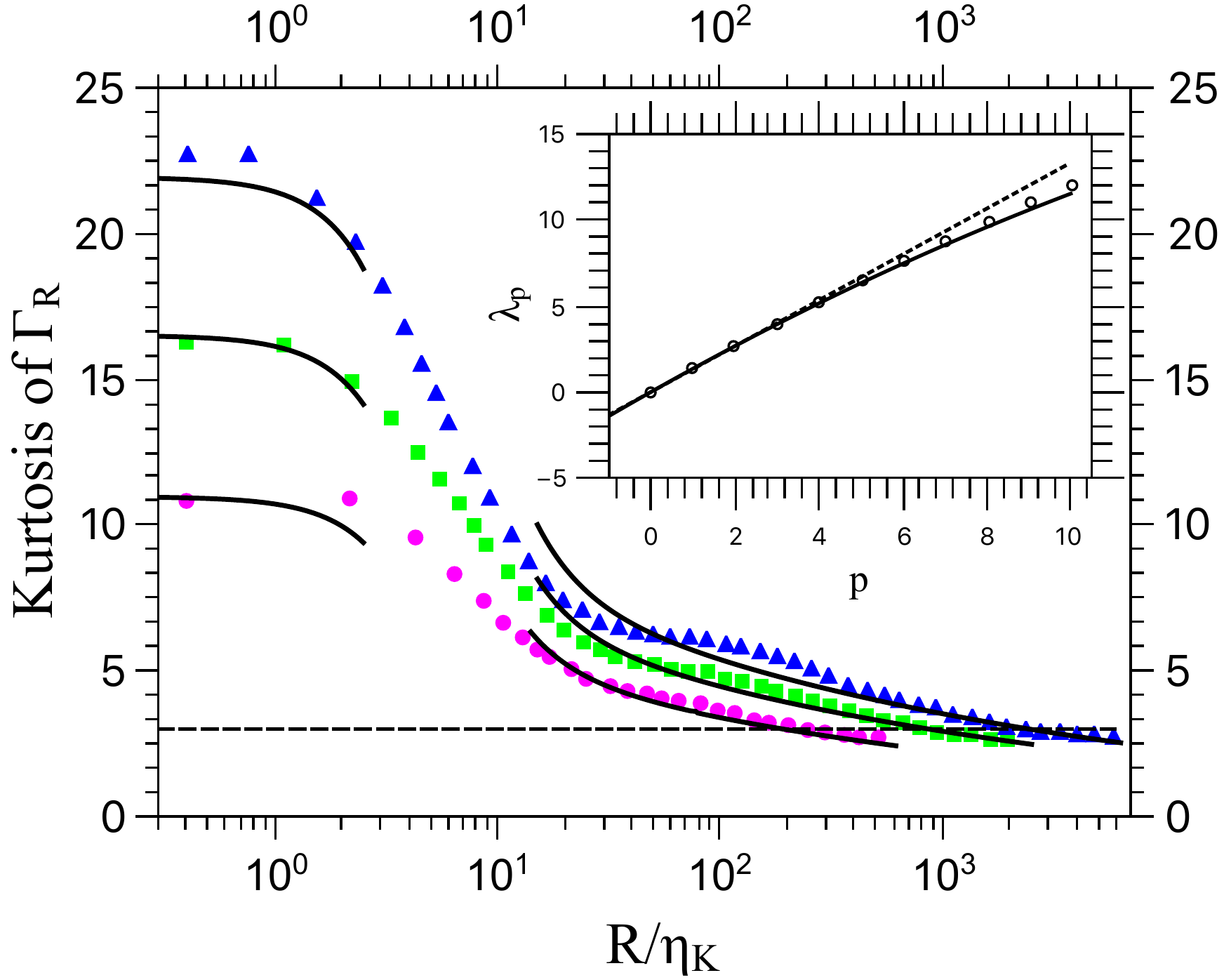}
\vspace{0.0cm}
\caption{Numerical kurtoses of circulation for $R_\lambda = 240$ ($\Circle$), $650$ ($\square$) and $1300$ ($\triangle$) from \cite{Iyer_etal}. Solid lines represent the asymptotic expansions, Eqs. (\ref{k4lowR}), with $a=3.3$, and (\ref{k4largeR}), with $b=2.5$, that hold at dissipative ($R \ll \eta_K$) and larger inertial range scales ($R \gg \eta_K$), respectively. The dashed line indicates, for reference, the kurtosis of a Gaussian distribution. The inset compares the numerical scaling exponents $\lambda_p$ of circulation moments of order $p$ evaluated at $R_\lambda = 1300$ \cite{Iyer_etal} (open circles) with the predicted values computed from Eq. (\ref{lambdap}) (solid line). The straight dotted line gives the K41 scaling, $\lambda_p = 4p/3$.} 
\label{kurt_expns}
\end{figure}

Moving, now, to the opposite asymptotic region {\hbox{$\bar R \gg 1$}}, we get, keeping all the subdominant terms, the circulation kurtosis
\be
\frac{\langle \Gamma_R^4 \rangle}{\langle \Gamma_R^2 \rangle^2} = \frac{\langle \xi_R^4 \rangle}{\langle \xi_R^2 \rangle^2} 
\frac{\sum_{n=1}^6 A_n \bar R^{\frac{4 + 2n}{3}}}{\sum_{n=1}^2 B_n \bar R^{\frac{4+2n}{3}}} \ , \ \label{k4largeR}
\ee
where the exactly computed coefficients $A_n's$ and $B_n's$ are
\bea
&&A_1 = 3 \delta \pi + 6 c^2 \delta^2 G \ , \ A_2 = 12 c \delta^2 E \ , \  \nonumber \\
&&A_3 = 12 c^2 \delta^3 F \ , \ A_4 = 3 \delta^2 \pi^2 \ , \  \nonumber \\
&&A_5 = 6c \delta^3 \pi E \ , \ A_6 = 3c^2\delta^4 E^2 \ , \  \nonumber \\
&&B_1 = \delta \pi \ , \ B_2 = c \delta^2 E \ , \ 
\eea
with 
\bea
&&c =  1/[2 \pi \Gamma(4/3)] \ , \ \delta = \bar \sigma \eta_K^2 \ , \ \nonumber \\
&&E = 4 \pi^{5/2} \Gamma(5/6) \Gamma(2/3) / [\Gamma(4/3) \Gamma(7/3)] \ , \ \nonumber \\
&&F = 0.33 \times (2 \pi)^5 \ , \ G = 2^{4/3} \pi^4 \Gamma(2/3) \ . \ 
\eea
It is worth emphasizing that $b$ is the only free modeling parameter in Eq. (\ref{k4largeR}). General moments of circulation can be computed for $\bar R \gg 1$, as well. At leading order,
we find
$
\langle \Gamma_R^p \rangle \sim R^{\frac{4p}{3}} \langle \xi_R^p \rangle \sim R^{\lambda_p} 
$,
where
\be
\lambda_p = \frac{4p}{3} - \frac{\mu}{8} p (p-2) \ . \ \label{lambdap}
\ee
The model predictions based on Eqs. (\ref{k4lowR}), (\ref{k4largeR}), and (\ref{lambdap}) lead to very suggestive
comparisons to numerical results, as shown in Fig. \ref{kurt_expns}. Eq. (\ref{lambdap}) implies that 
$d \lambda_p / dp = 4/3 + \mu/4 = 1.376 \pm 0.002$ at $p=0$. This result is in striking agreement with the value $1.367 \pm 0.009$ obtained from accurate numerical simulations \cite{Iyer_etal}.

\begin{figure}[ht]
\hspace{0.0cm} \includegraphics[width=0.48\textwidth]{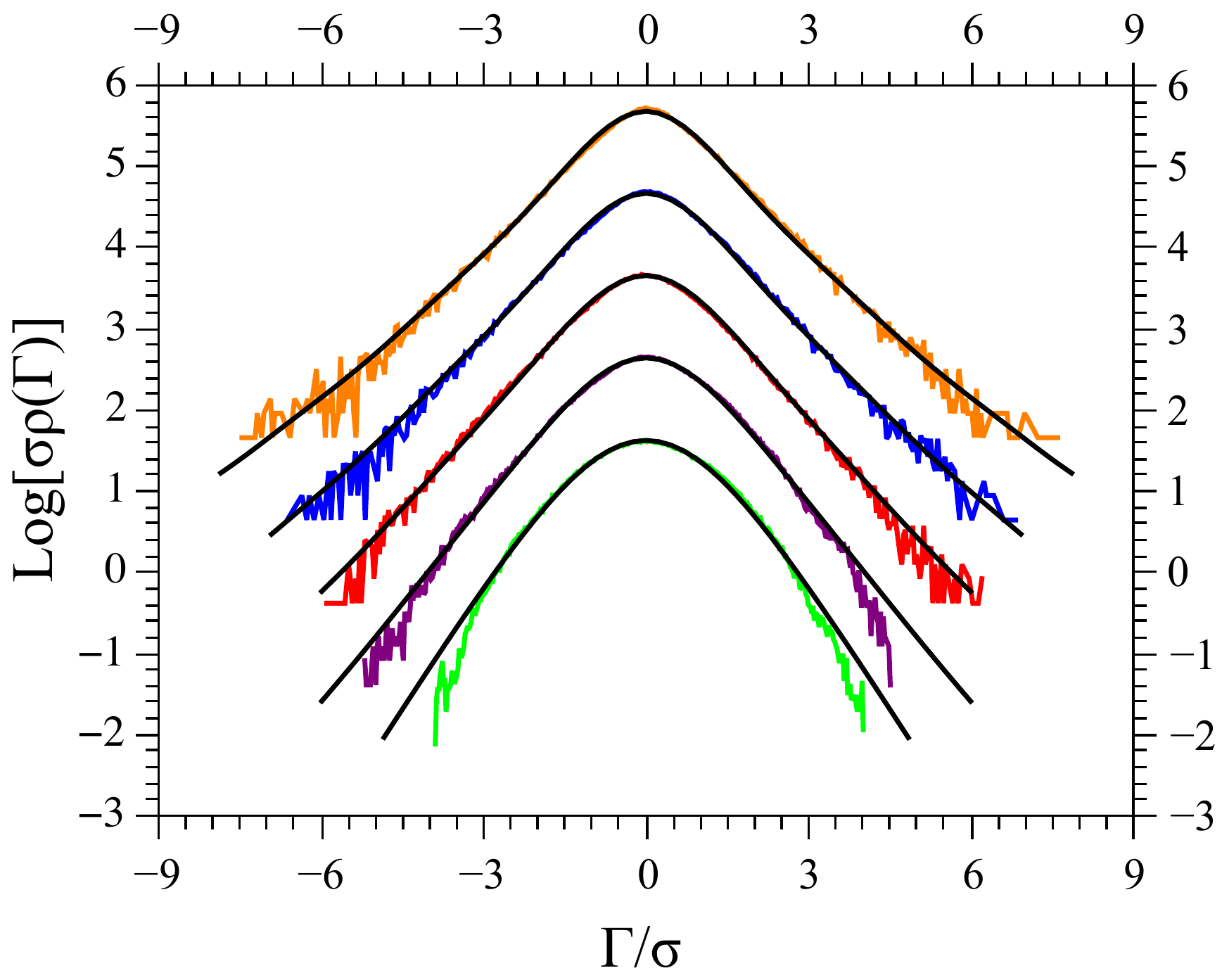}
\vspace{0.0cm}
\caption{Comparisons between standardized numerical cPDFs (vertically displaced to ease visualization) for {\hbox{$R/\eta_K = 16$, $32$, $64$, $128$ and $256$}} (from the top to the bottom) obtained at $Re_\lambda = 418$ from the Johns Hopkins University turbulence database \cite{JHTD}, and the ones evaluated from the theoretical result (solid black lines) given by Eq. (\ref{cPDF}). The cPDF tails are in general slightly subestimated due to finite size ensemble effects, with more pronounced drops taking place at larger circulation contours.}
\label{cPDFs}
\end{figure}

The present approach, furthermore, allows us to derive closed analytical expressions for the circulation probability distribution functions (cPDFs), $\rho_R(\Gamma)$, for $\bar R \gg 1$.
Using (\ref{circ}) and (\ref{omegaR}), and representing, for convenience, the circulation in units of $2 \pi \eta^2 \sqrt{\epsilon_0 / (3 \nu)}$, the associated characteristic function is written as the triple expectation value
\bea
&&Z(\zeta) =  \langle  \langle  \langle \exp \left [ i \zeta \Gamma_R \right ]  \rangle \rangle \rangle_{( \tilde \omega, \xi_R, \sigma ) } \nonumber \\
&&=  \int_0^\infty d \xi f_R(\xi) \langle \exp \left [ -\frac{1}{2} \zeta^2 \xi^2 \Omega \right ] \rangle_\sigma \ , \ \label{Zzeta}
\eea
where $f_R(\xi)$ is the lognormal probability distribution function for the random variable $\xi_R$, defined from (\ref{xiX}) and (\ref{Xxi}), and 
\be
\Omega \equiv \int_\mathcal{D} d^2 \bfr \int_\mathcal{D} d^2 \bfr' \left \langle \tilde \omega (\bfr) \tilde \omega(\bfr') 
\right \rangle
\sigma(\bfr) \sigma(\bfr') \  . \ 
\ee
It is not difficult to show that the variance of $\Omega$ becomes very small compared to $\bar \Omega^2$ in the region $\bar R \gg 1$. Therefore,
averaging over the $\sigma(\bfr)$ fields is effectively equivalent to replacing $\Omega$ by its mean value
$\bar \Omega \propto R^{\frac{8}{3}}$ in (\ref{Zzeta}). Performing the Fourier transform of (\ref{Zzeta}), we get
\be
\rho_R(\Gamma) = \frac{1}{\sqrt{2 \pi \bar \Omega}}  \int_0^\infty d \xi \frac{1}{\xi} f_R ( \xi) 
\exp \left ( - \frac{\Gamma^2}{2 \xi^2 \bar \Omega} \right ) \ , \ \label{cPDF}
\ee
which is remarkably analogous to the Castaing et al. modeling form of velocity increment PDFs \cite{castaing_etal,beck,prf19}, with $\xi$ taking the place of the energy dissipation rate $\epsilon$. Additional analysis shows that for $1 \ll \bar R \ll R_\lambda^{3/2}$, the cPDF just obtained can be recast as a function of $\Gamma \exp(\bar X_R)/\sqrt{\bar \Omega} \sim \Gamma / R^h$ around its central core, where $h = 4/3 + \mu /4 = 1.376$, in agreement with the observations of Ref. \cite{Iyer_etal}. Numerical cPDFs computed from the Johns Hopkins University turbulence data base \cite{JHTD} are closely reproduced by Eq. (\ref{cPDF}), as indicated in Fig. \ref{cPDFs}.

To summarize, we have been able to address several important statistical features of turbulent circulation, relying on the fusion of structural concepts - the picture of a turbulent flow as a system of sparse vortex tubes - with the long known OK62 phenomenological approach to intermittency. The vortex gas modeling introduced here throws light on the dependence of the circulation kurtosis with the Reynolds number and probing scale, the scaling exponents of the circulation moments, the detailed shape of cPDFs and their related collapsing exponent $h \simeq 1.4$.

We have not touched on further issues as the area law for the cPDFs, bifractality, and the tendency for the formation of a circulation kurtosis plateau in the deep inertial range region, as observed at higher Reynolds numbers \cite{Iyer_etal}. These points are likely to be more conveniently approached by Monte Carlo simulations of the fields $\tilde \omega (\bfr)$, $\xi (\bfr)$, and $\sigma (\bfr)$, under a variety of model definitions. The formulation of a bridge between the languages of vortex gas modeling and multifractality is also an exciting problem, which surely deserves close attention in future investigations.

\section{Acknowledgments}
This work has been partially supported by CNPq.


\begin{references}


\bibitem{wilson} K.G. Wilson, Phys. Rev. D {\bf{10}}, 2445 (1974).

\bibitem{migdal1} A.A. Migdal, Phys. Rep. {\bf{102}}, 199 (1983).

\bibitem{migdal2} A.A. Migdal, Int. J. Mod. Phys. A {\bf{9}}, 1197 (1994).

\bibitem{migdal3} A.A. Migdal {\it{Universal Area Law in Turbulence}}, https://arxiv.org/abs/1903.08613v3.

\bibitem{Iyer_etal} K. P. Iyer, K. R. Sreenivasan, and P. K. Yeung, Phys. Rev. X {\bf{9}}, 041006 (2019).
\bibitem{migdal4} A.A. Migdal {\it{Probability Distribution of Velocity Circulation in Three Dimensional Turbulence}}, https://arxiv.org/abs/2006.12008.

\bibitem{orszag_etal} Z.-S. She, E. Jackson, and S.A. Orszag, Nature {\bf{344}}, 226 (1990).

\bibitem{farge_etal} M. Farge, G. Pellegrino, and K. Schneider, Phys. Rev. Lett. {\bf{87}}, 054501 (2001).

\bibitem{kaneda_etal} T. Ishihara, T. Gotoh, and Y. Kaneda, Annu. Rev. Fluid Mech. {\bf{41}}, 165 (2009).

\bibitem{K41} A.N. Kolmogorov, Dokl. Akad. Nauk SSSR {\bf{30}}, 301 (1941); 
Proc. R. Soc. Lond. A {\bf{434}}, 9 (1991).

\bibitem{bolt} E. Bolthausen, Ann. Probab. {\bf{10}}, 1047 (1982).

\bibitem{O62} A. M. Obukhov, J. Fluid Mech. {\bf{13}}, 77 (1962).

\bibitem{K62} A. N. Kolmogorov, J. Fluid Mech. {\bf{13}}, 82 (1962).

\bibitem{frisch} U. Frisch, {\it{Turbulence}}, Cambridge University Press (1995).

\bibitem{ro_va} R. Robert and V. Vargas, Ann. Probab. {\bf{38}}, 605 (2010).
\newpage

\bibitem{jimenezetal93} J. Jimenez, A. A. Wray, P. G. Saffman and R. S. Rogallo, J. Fluid Mech., {\bf 255}, 65 (1993).

\bibitem{wyn_tenne} J.C. Wyngaard and H. Tennekes, Phys. Fluids {\bf{13}}, 1962 (1970).

\bibitem{kholmy_etal} M. Kholmyansky, L. Moriconi, R.M. Pereira, and A. Tsinober, Phys. Rev. E {\bf{80}}, 036311 (2009).

\bibitem{pereira_etal} R.M. Pereira, L. Moriconi, and L. Chevillard, J. Fluid Mech. {\bf{839}}, 430 (2018).

\bibitem{batchelor} G.K. Batchelor, {\it{The Theory of Homogeneous Turbulence}}, Cambridge University Press (1953).

\bibitem{comment} A quick check shows that if $\Delta N$ is defined as the integral of $\phi(\bfr)$ over any bounded domain, then the moments of $\Delta N$ satisfy, up to fourth-order, the usual Poisson relations.

\bibitem{tang_etal} S.L. Tang, R.A. Antonia, L. Djenidi, and Y. Zhou, J. Fluid Mech. {\bf{891}}, A26 (2020).

\bibitem{JHTD} Y. Li, E. Perlman, M. Wan, Y. Yang, R. Burns, C. Meneveau, R. Burns, S. Chen, A. Szalay and G. Eyink, J. Turbul. {\bf 9} N31 (2008).

\bibitem{castaing_etal} B. Castaing, Y. Gagne, and E. J. Hoppfinger, Physica D {\bf{46}}, 177 (1990).

\bibitem{beck} C. Beck, Phys. Rev. Lett. {\bf 87}, 180601 (2001).

\bibitem{prf19} W. Sosa-Correa, R. M. Pereira, A. M. S. Macêdo, E. P. Raposo, D. S. P. Salazar and G. L. Vasconcelos, Phys. Rev. Fluids {\bf{4}}, 064602 (2019).

\end{references}
\end{document}